\documentclass[aps,prc,preprint,showpacs,floatfix]{revtex4-1}
\usepackage[dvips]{color}
\usepackage{graphicx}
\usepackage{bm}

\begin{document}

\title{Linearly increasing radius of the light fragment during the
spontaneous fission of $^{282}$Cn}

\author{D. N. Poenaru$^*$ and R. A. Gherghescu}
\email[]{poenaru@fias.uni-frankfurt.de}
\affiliation{
Horia Hulubei National Institute of Physics and Nuclear
Engineering (IFIN-HH), \\P.O. Box MG-6, RO-077125 Bucharest-Magurele,
Romania and\\ Frankfurt Institute for Advanced Studies, Johann Wolfgang
Goethe University, Ruth-Moufang-Str. 1, D-60438 Frankfurt am Main, Germany}

\date{ }

\begin{abstract}

In a previous article published in Phys.  Rev.  C 94 (2016) 014309 we have
shown for the first time that the best dynamical trajectory during the
deformation toward fission of the superheavy nucleus $^{286}$Fl is a
linearly increasing radius of the light fragment, $R_2$.  This
macroscopic-microscopic result reminds us about the $\alpha $ or cluster
preformation at the nuclear surface, assumed already in 1928, and proved
microscopically many times.  This time we give more detailed arguments for
the neighboring nucleus $^{282}$Cn.  Also similar figures are presented for
heavy nuclei $^{240}$Pu and $^{252}$Cf.  The deep minimum of total
deformation energy near the surface is shown for the first time as a strong
argument for cluster preformation.

\end{abstract}

\pacs{25.85.Ca, 24.75.+i, 21.10.Tg, 27.90.+b}

\maketitle

\section{Introduction}
\label{sec:1}

The most important decay modes of superheavy nuclei are mainly
$\alpha$~decay and spontaneous fission
\cite{khu14prl,ham13arnps,oga11ra,hof11ra,due10prl,mor07jpsjb,oga07jpg,hof00rmp,aud12cpc1}.
Among the many theoretical papers in this field one should mention
\cite{sob11ra,wan15prc,sob16jpg,qic16rp,qia16prc} and
\cite{sta13prc,war11prc,smo97pr,smo95pr,bao15jpg,bao13np,san10npa,xu08prc}. 
For atomic numbers larger than 121 cluster decay \cite{enc95,p195b96} may
compete as well \cite{p315prc12,p309prl11}.

In 1928 G. Gamow \cite{gam28zp} as well as R.W. Gurney and E.U. Condon
\cite{gur28n} gave the first explanation of $\alpha $~decay based on quantum
mechanical tunneling of a preformed particle at the nuclear surface. The
microscopic theory had been developed by many scientists, e.g.
\cite{tho54ptp,lan60rmp,man64arns,ton79np,fli85zp,var92prl,ble96mb}.
It was also extended to explain cluster decays
\cite{ble96mb,lov98pr,del10b}. Simple relationships are also very useful
\cite{p83jpl80,wan15prc}.

In our paper mentioned in the abstract \cite{p348prc16} we reported results
obtained within macroscopic-microscopic method \cite{str67np} using cranking
inertia \cite{ing54pr,bra72rmp} and the best two-center shell model
\cite{ghe03prc} in the plane of two independent variables $(R,\eta)$, where
$R$ is the separation distance of the fragments and $\eta =(A_1 - A_2)/A$ is
the mass asymmetry with $A, A_1, A_2$ the mass numbers of the parent and
nuclear fragments.  Phenomenological deformation energy, $E_{Y+E}$, was
given by Yukawa-plus-exponential model \cite{kra79pr}, and the shell plus
pairing corrections, $\delta E = \delta U + \delta P$ are based on the
asymmetric two center shell model (ATCSM) \cite{ghe03prc}.  This time we
give more detailed arguments for the neighboring nucleus $^{282}$Cn.  Also
similar figures are presented for heavy nuclei $^{240}$Pu and $^{252}$Cf. 
The deep minimum of total deformation energy near the surface is shown for
the first time as a strong argument for cluster preformation.

\section{Model}

An outline of the model was presented previously \cite{p348prc16}. We repeat
few lines in this section.  The parent $^A Z$ is split in two fragments: the
light, $^{A_2}Z_2$, and the heavy one, $^{A_1}Z_1$ with conservation of
hadron numbers $A=A_1+A_2$ and $Z=Z_1+Z_2$.  The corresponding radii are
given by $R_0=r_0 A^{1/3}$, $R_{2f}=r_0 A_2^{1/3}$, and $R_{1f}=r_0
A_1^{1/3}$.  The separation distance of the fragments is initially $R_i =
R_0$ and at the touching point $R_t = R_{1f} + R_{2f}$ with $r_0=1.16$~fm.

The geometry for linearly increasing $R_2$ from 0 to $R_{2f}=R_e$ is defined
by:
\begin{equation}
R_2 = R_{2f}\frac{R-R_i}{R_t-R_i}
\end{equation}

According to the macroscopic-microscopic method the total deformation energy
contains the Yukawa-plus-exponential (Y+EM) and the shell plus pairing
corrections
\begin{equation} 
E_{def} = E_{Y+E} + \delta E
\end{equation}
In units of $\hbar \omega_0^0 = 41 A^{-1/3}$ the shell corrections are
calculated with the Strutinsky procedure as a sum of protons and neutrons
contributions
\begin{equation}
\delta u = \delta u_p + \delta u_n
\end{equation}
By solving the BCS \cite{bar57pr} system of two equations with two
unknowns, we find the Fermi energy, $\lambda$, and the pairing gap $\Delta$.
The total pairing corrections are given by
\begin{equation}
\delta p = \delta p_p + \delta p_n
\end{equation}
and finally the total shell plus pairing corrections in MeV
\begin{equation}
\delta E = \delta U + \delta P 
\end{equation}
The inertia tensor \cite{bra72rmp} is given by
\begin{equation}
B_{ij} =2\hbar^2\sum_{\nu \mu} \frac{\langle \nu|\partial H/\partial
\beta_i|\mu \rangle \langle \mu|\partial H/\partial \beta_j|\nu
\rangle}{(E_\nu +E_\mu)^3}(u_\nu v_\mu +u_\mu v_\nu)^2  
\label{eq3}
\end{equation}
where $H$ is the single-particle Hamiltonian allowing to determine the
energy levels and the wave functions $|\nu \rangle$, $u_\nu^2$, $v_\nu^2$
are the BCS occupation probabilities, $E_\nu$ is the quasiparticle energy,
and $\beta_i, \beta_j$ are the independent shape coordinates.
For spherical fragments with $R, R_2$ deformation parameters the cranking
inertia symmetrical tensor will have three components, hence the scalar
\begin{equation}
B(R) =  B_{R_2R_2}\left( \frac{dR_2}{dR} \right)^2+
2B_{R_2 R}\frac{dR_2}{dR} + B_{RR}  = B_{22} + B_{21} + B_{11}
\label{br}
\end{equation}

When we find the least action trajectory in the plane $(R,R_2)$ we need to
calculate the three components $B_{22}, B_{21}, B_{11}$ in every point of a
grid of 66$\times$24 (for graphics) or 412$\times$24 (for the real
calculation) for 66 or 412 values of $(R-R_i)/(R_t-R_i)$ and 24 values of
$\eta = (A_1-A_2)/A$ or $R_{2f}$.

\section{Results}

\begin{figure}[ht]
\begin{center}
\includegraphics[width=10cm]{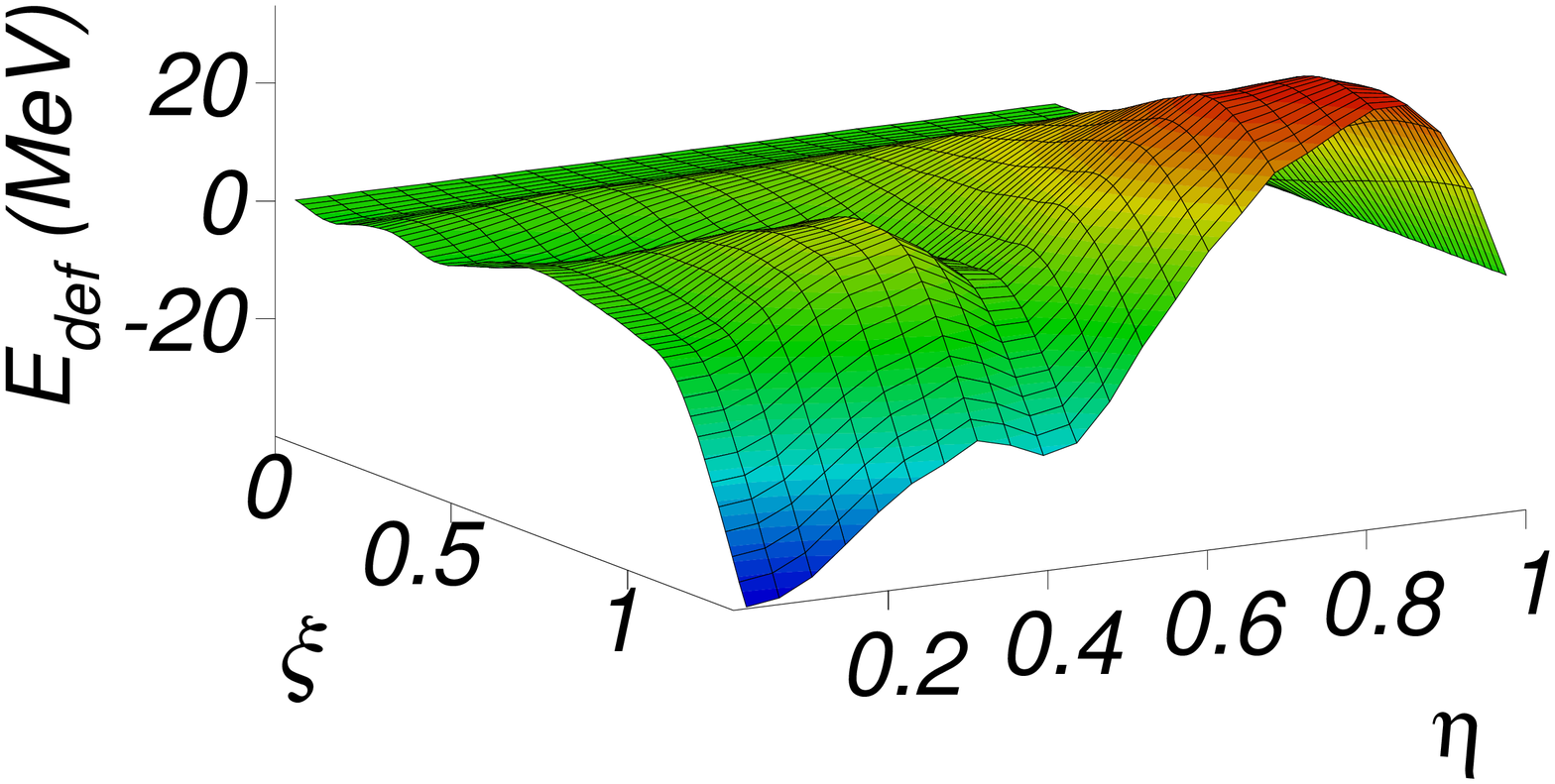} 
\vspace*{-0.9cm}

\includegraphics[width=10cm]{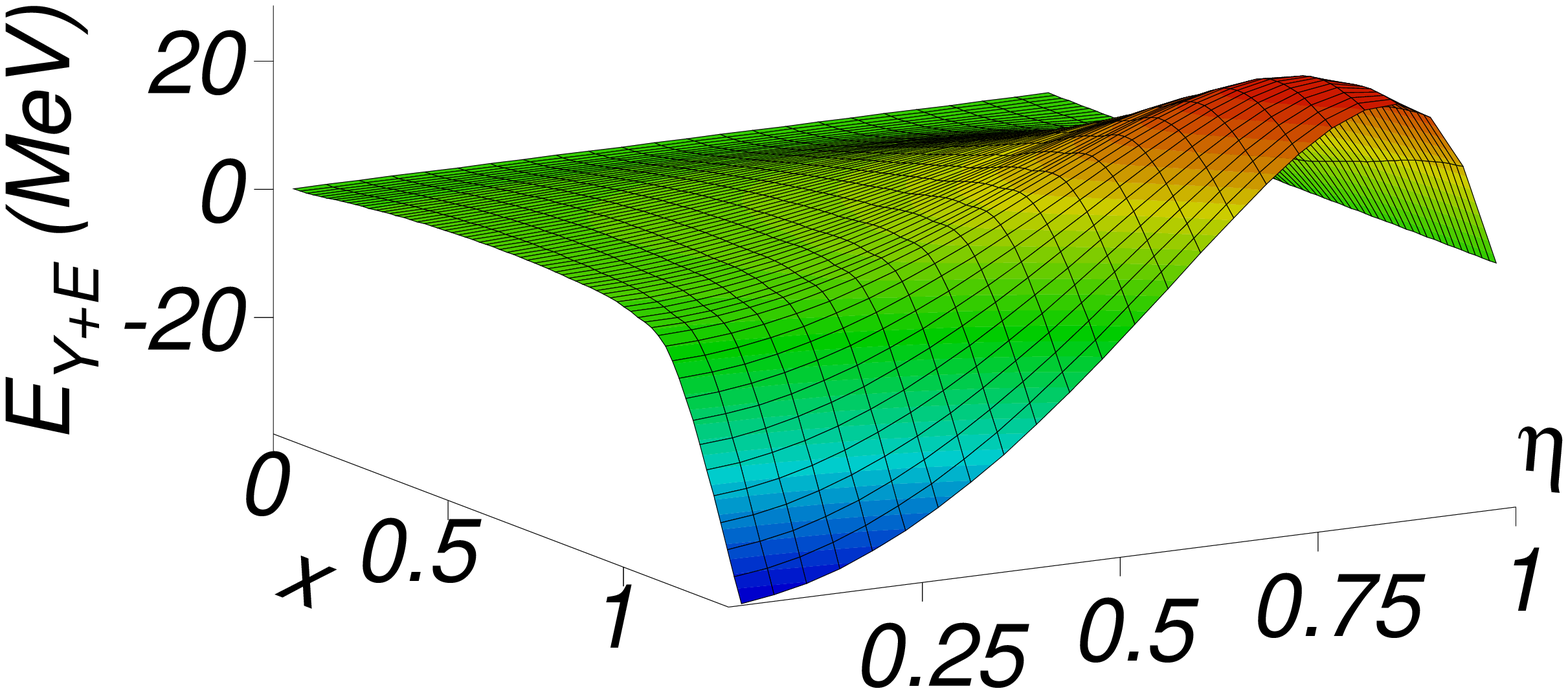} 

\end{center}
\caption{(Color online) PES of $^{282}$Cn vs $(R-R_i)/(R_t - R_i) \geq 0$
and $\eta = (A_1-A_2)/(A_1+A_2)$.  Y+EM (bottom),  and total deformation 
energy (top). $R_2 = $constant.
\label{pescn}} 
\end{figure}

\begin{figure}[ht]
\begin{center}
\includegraphics[width=10cm]{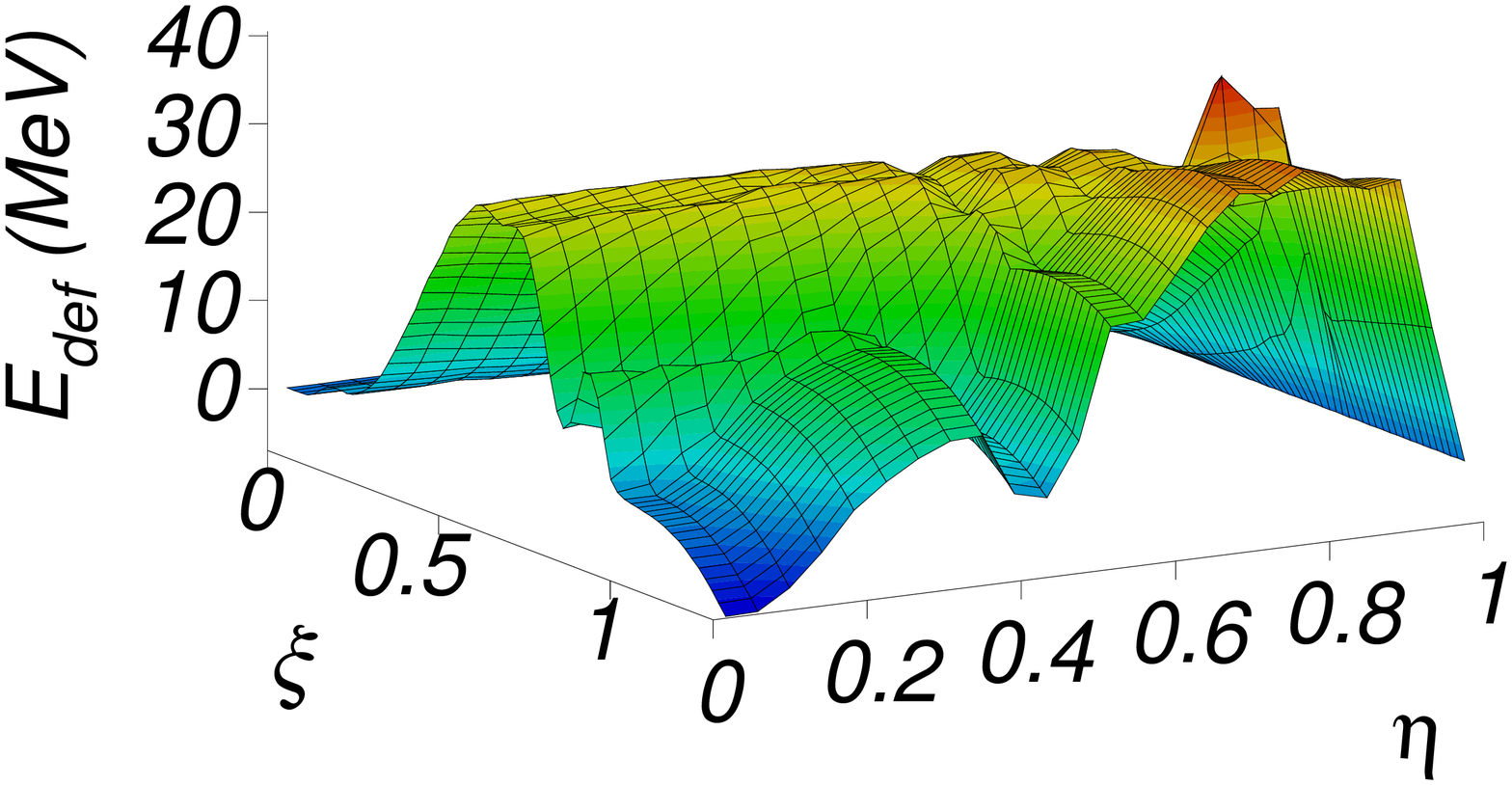} 
\vspace*{-0.9cm}

\includegraphics[width=10cm]{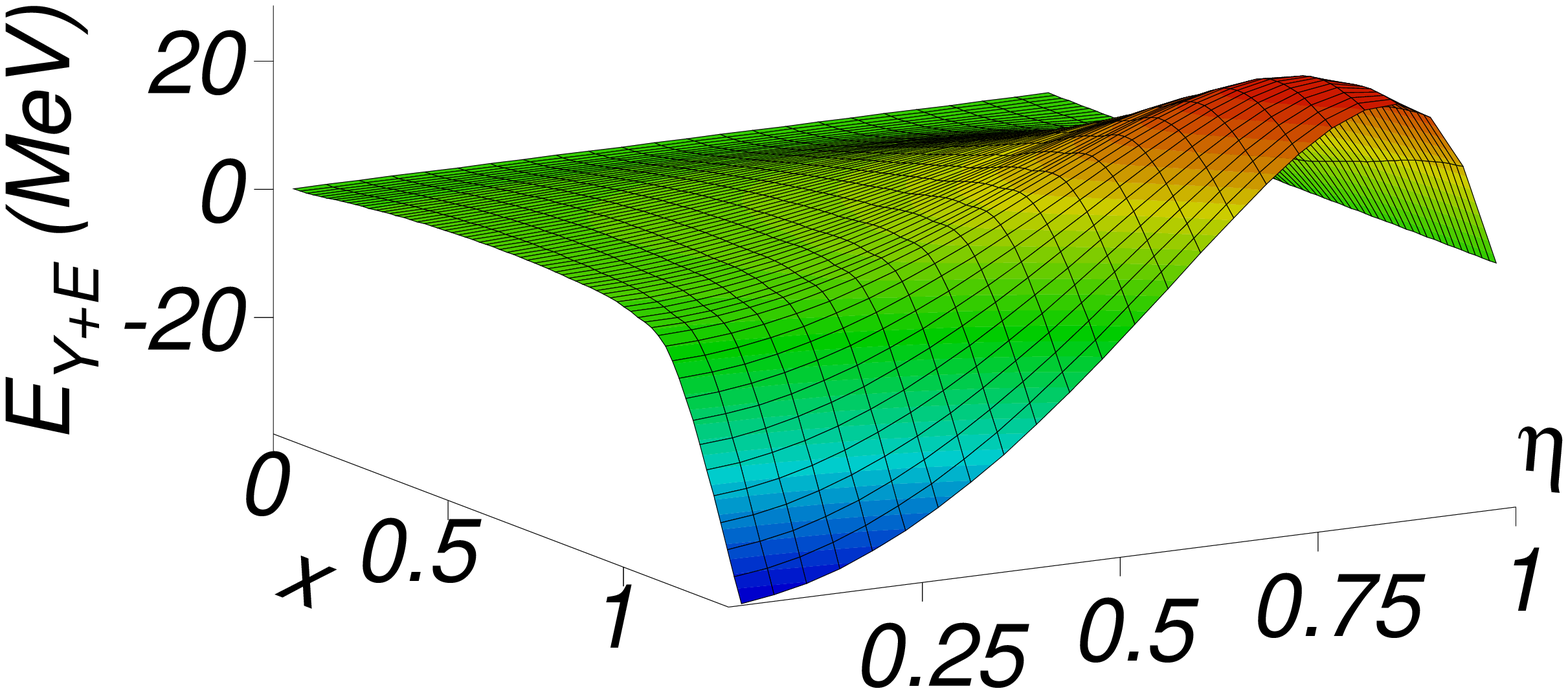} 

\end{center}
\caption{(Color online) PES of $^{282}$Cn vs $(R-R_i)/(R_t - R_i) \geq 0$
and $\eta = (A_1-A_2)/(A_1+A_2)$.  Y+EM (bottom),  and total deformation 
energy (top). $R_2 $ linearly increasing with $R$.
\label{pescnl}} 
\end{figure}

\begin{figure}
\centerline{\includegraphics[width=10cm]{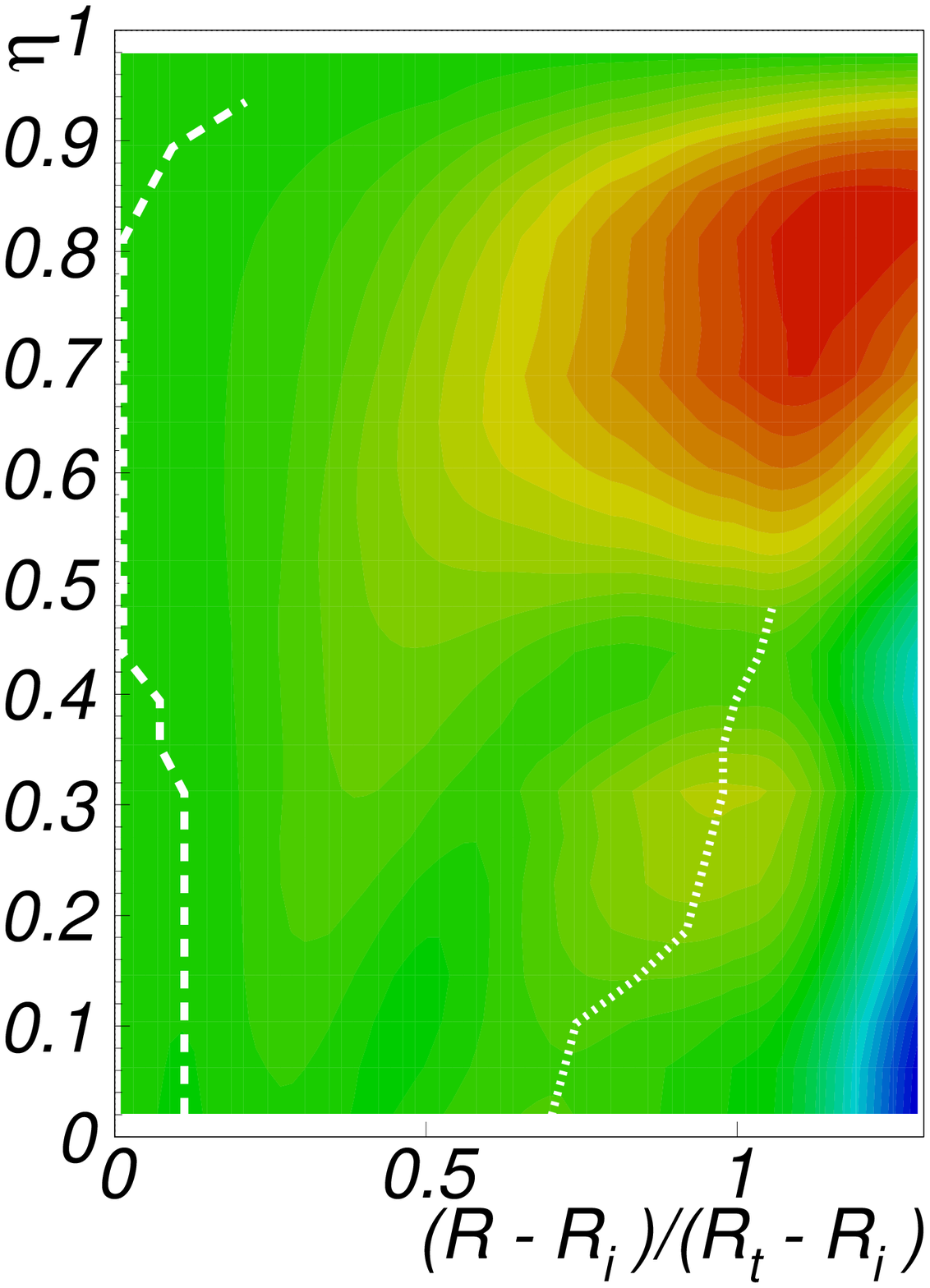}} 
\vspace*{-1cm} 
\caption{(Color online) Contour plot of deformation energy of $^{282}$Cn
shown as a PES in the upper panel of Fig.~\ref{pescn}.  The first and
second minima of deformation energy at every value of mass asymmetry are
plotted with dashed and dotted white lines. $R_2 = $constant.
\label{cnc}} 
\end{figure}

\begin{figure}
\centerline{\includegraphics[width=10cm]{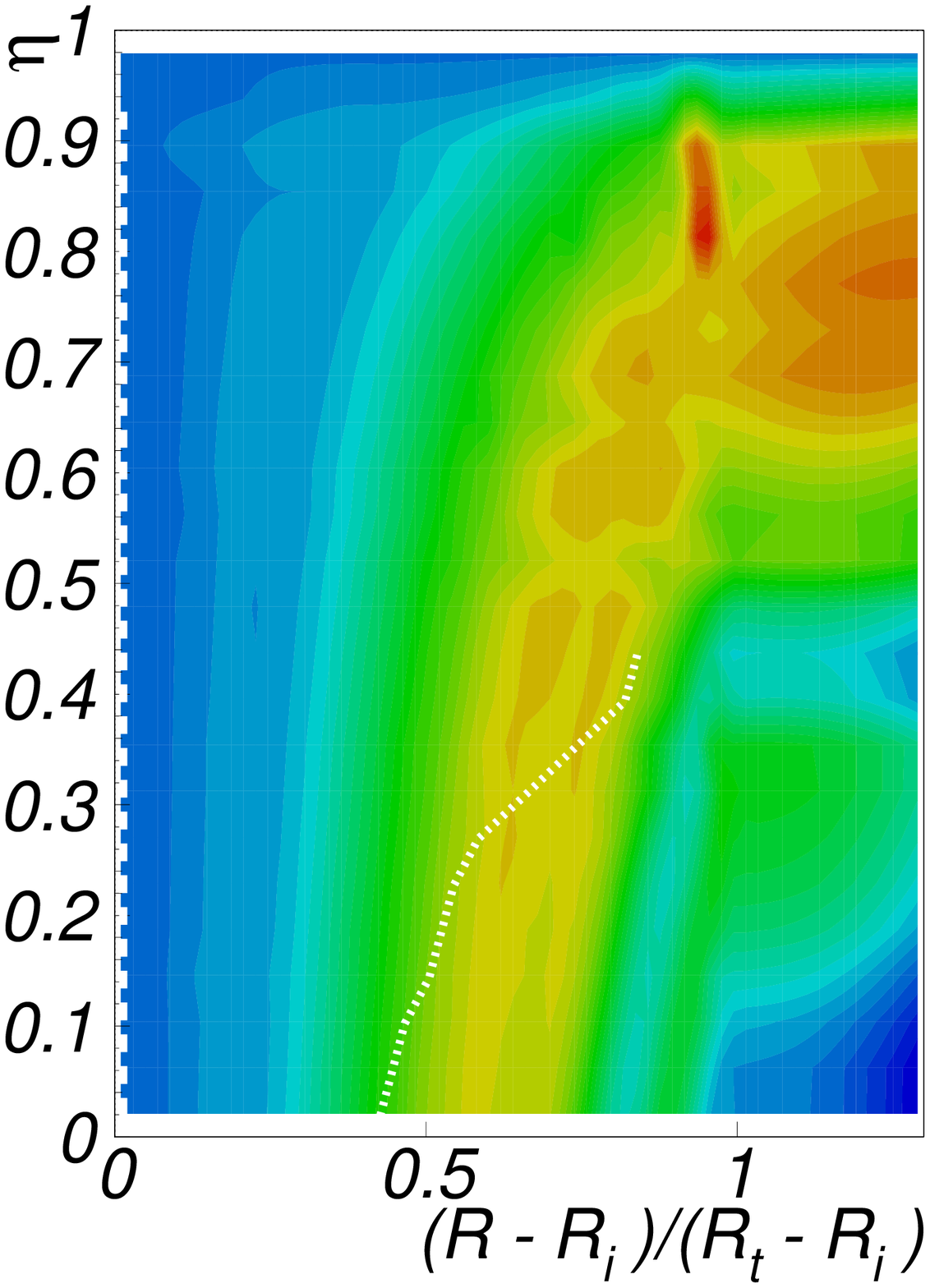}} 
\vspace*{-1cm} 
\caption{(Color online) Contour plot of deformation energy of $^{282}$Cn
shown as a PES in the upper panel of Fig.~\ref{pescnl}.  The first and
second minima of deformation energy at every value of mass asymmetry are
plotted with dashed and dotted white lines. $R_2 $ linearly increasing with
$R$.
\label{cncl}} 
\end{figure}

Potential energy surfaces (PES) for spontaneous fission of $^{282}$Cn are
shown in figures~\ref{pescn} and~\ref{pescnl} for constant radius, $R_2$, of
the light fragment and linearly increasing one, respectively.  The
corresponding contour plots are given in figures~\ref{cnc} and~\ref{cncl},
where the first and second minima of deformation energy at every value of
mass asymmetry are plotted with dashed and dotted white lines.  More details
are given in the two tables~\ref{tab} and~\ref{tabl}.  Also, the position
and value of maximum Y+EM model deformation energy versus mass asymmetry,
$\eta $, for fission of $^{282}$Cn with linearly increasing $R_2$ (top) and
constant $R_2$ (bottom) are shown in figure~\ref{eym}.   

\begin{figure}
\centerline{\includegraphics[width=10cm]{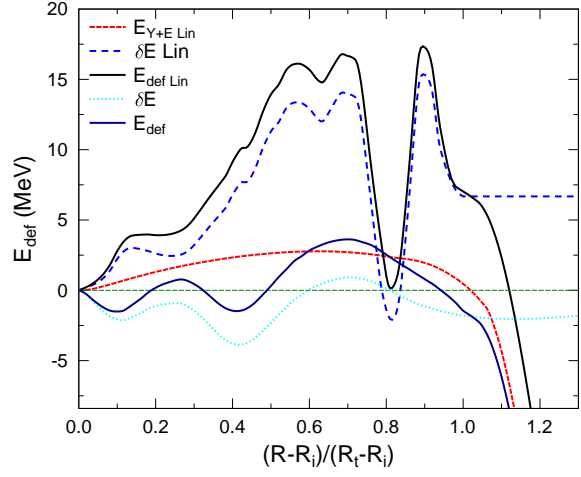}} 
\caption{(Color online) Comparison of deformation energies with respect to
spherical shapes for symmetrical fission of $^{282}$Cn with $R_2$~constant
and linearly increasing $R_2$ (Lin).
\label{cen}} 
\end{figure}

In figure \ref{cen} we compare the deformation energies with respect to
spherical shapes for symmetrical fission of $^{282}$Cn with $R_2$~constant
and linearly increasing $R_2$ (Lin).  One can see a relatively law
macroscopic energy $E_{Y+E Lin}$ on which the shell and pairing corrections,
$\delta E Lin$ deegs a rather deep minimum not far from the nuclear surface. 
A completely different ``classical'' two-humped barrier, $E_{def}$, may be
seen for $R_2 =$~constant.  
While the first minima in figure~\ref{cncl} and table~\ref{tabl}, are all
lying at $x=0$, this is true only for 10 mass asymetries out of 23 in
figure~\ref{cnc} and the table~\ref{tab}.

\begin{figure}
\centerline{\includegraphics[width=10cm]{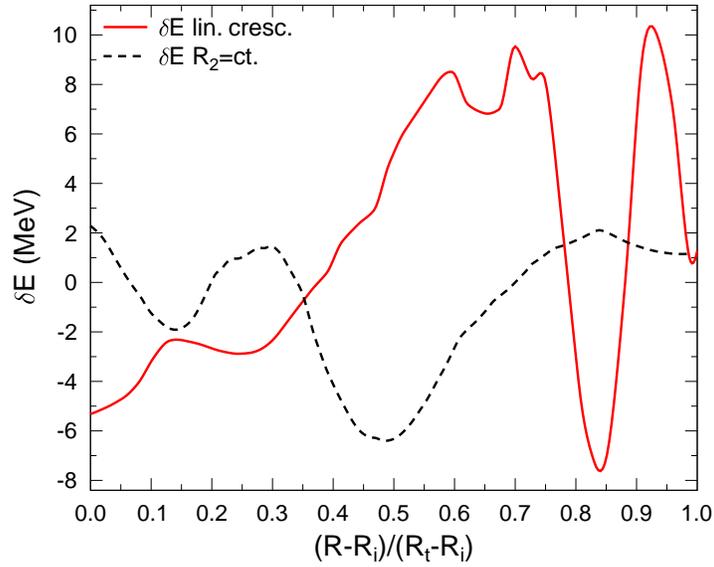}} 
\caption{(Color online) Comparison of absolute values of shell and pairing
correction energies for almost symmetrical fission of $^{282}$Cn with
$R_2$~constant (dashed line) and linearly increasing $R_2$ (solid line). 
\label{shpe2}} 
\end{figure}

\begin{figure} 
\centerline{\includegraphics[width=10cm]{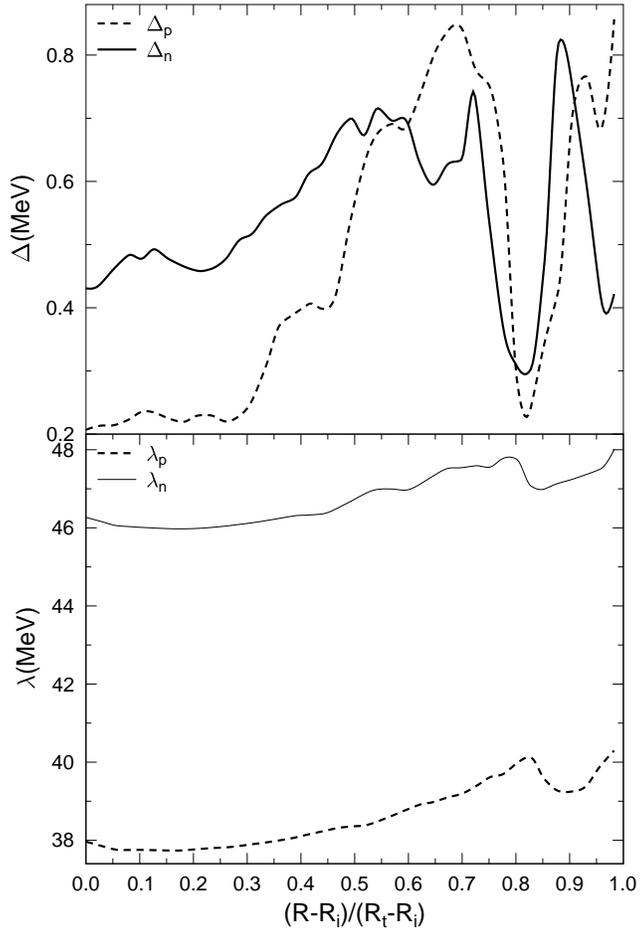}} 
\caption{(Color online) Solutions of BCS equations for symmetrical fission
of $^{282}$Cn with inearly increasing $R_2$: the gap for protons and
neutrons (top) and the Fermi energy for protons and neutrons (bottom).
\label{bcs}} 
\end{figure}

\begin{figure}
\centerline{\includegraphics[width=10cm]{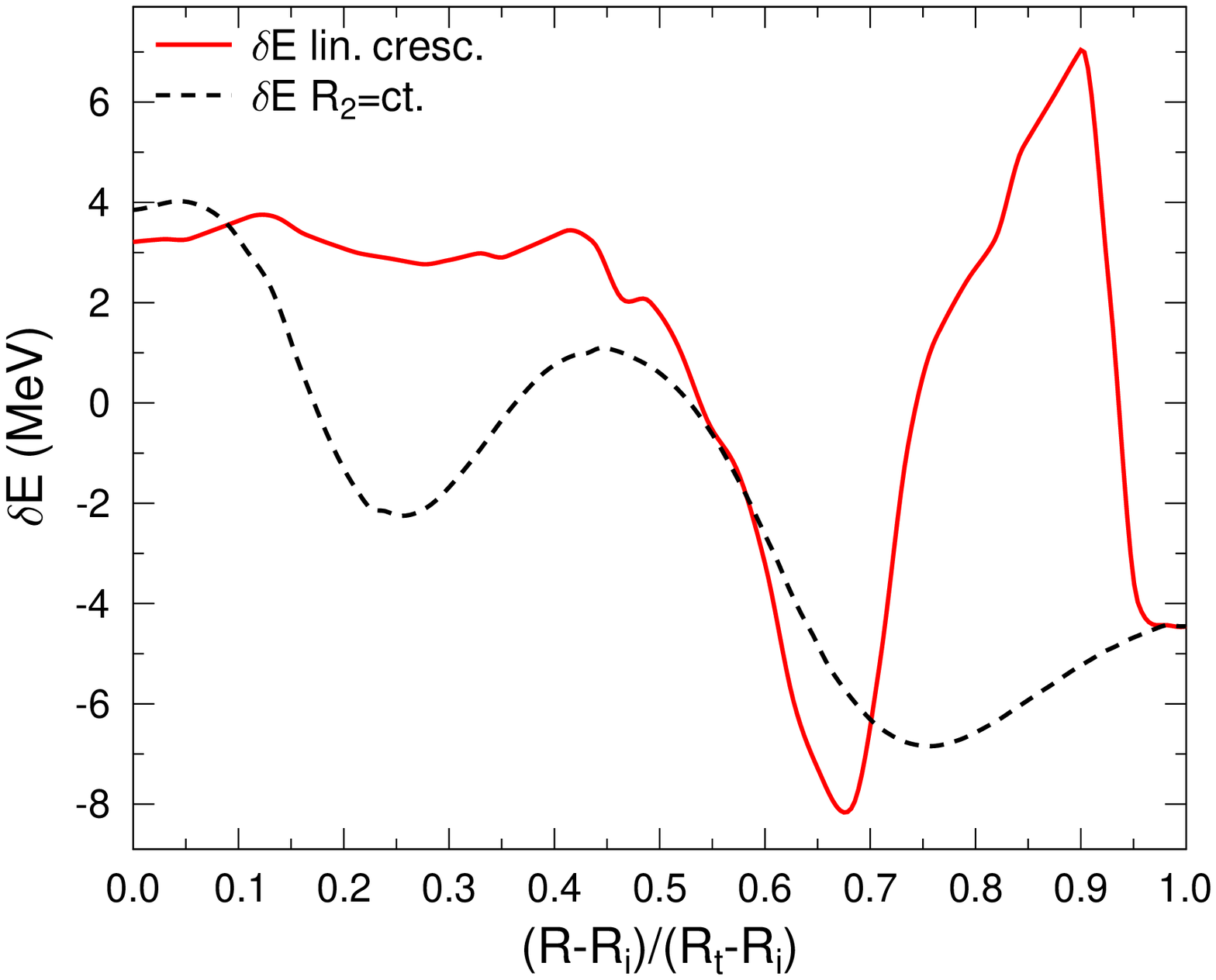}} 
\caption{(Color online) Comparison of shell plus pairing effects for fission 
of $^{240}$Pu with linearly increasing $R_2$ and constant $R_2$.
\label{pu00lc}} 
\end{figure}

\begin{figure}
\centerline{\includegraphics[width=10cm]{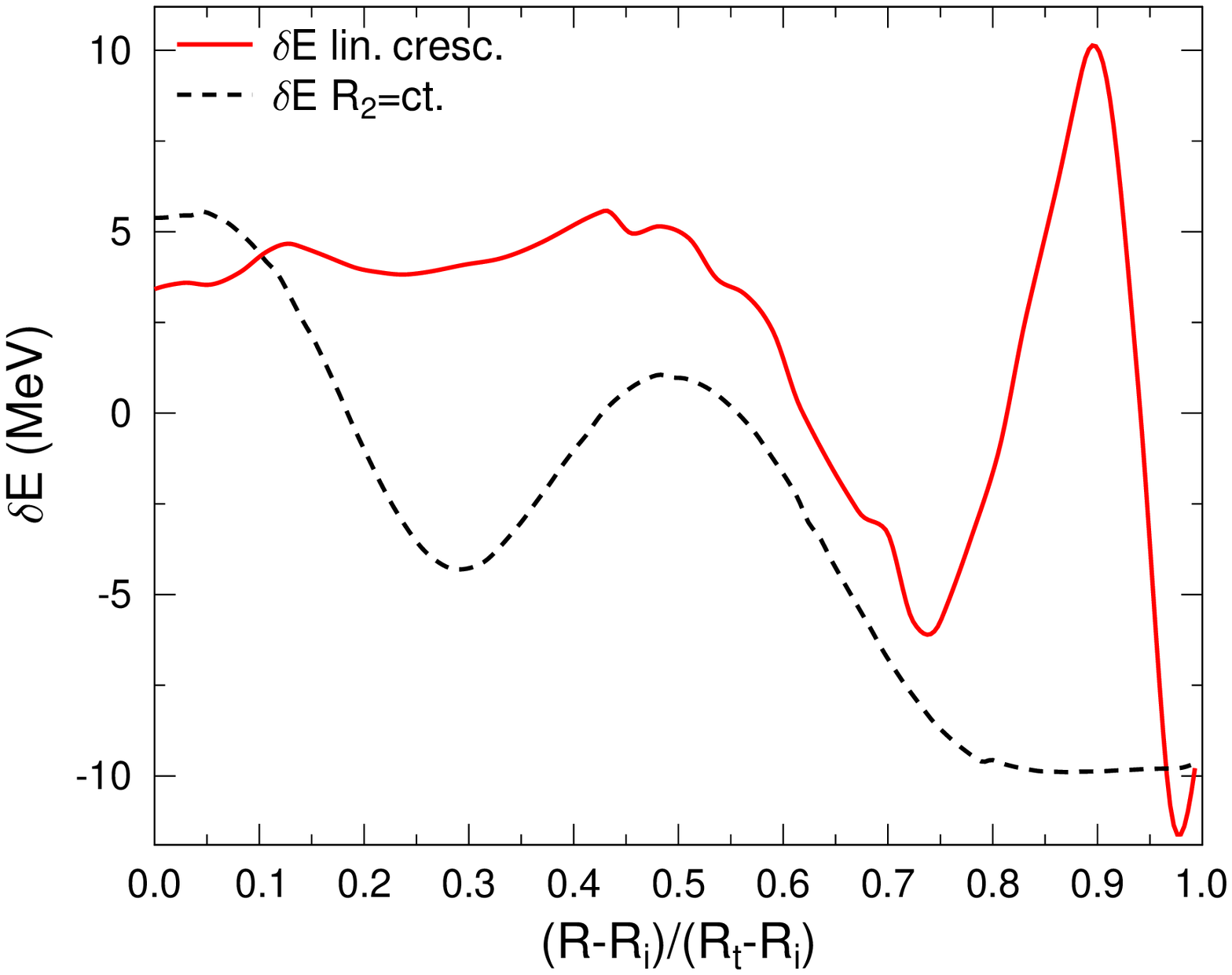}} 
\caption{(Color online) Comparison of shell plus pairing effects for fission 
of $^{252}$Cf with linearly increasing $R_2$ and constant $R_2$.
\label{cf00lc}} 
\end{figure}

\begin{figure}[ht]
\centerline{\includegraphics[width=8cm]{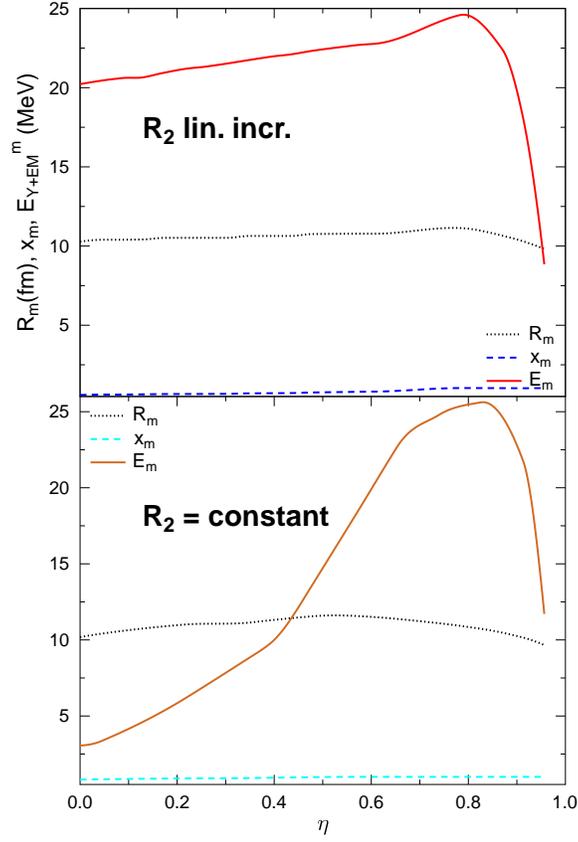}} 
\caption{(Color online) Position and value of maximum Y+EM model deformation
energy versus mass asymmetry, $\eta $, for fission of $^{282}$Cn with
linearly increasing $R_2$ (top) and constant $R_2$ (bottom).
\label{eym}} 
\end{figure}

We compare in figure \ref{shpe2} the absolute values of shell and pairing
correction energies for symmetrical fission of $^{282}$Cn with
$R_2$~constant (dashed line) and linearly increasing $R_2$ (solid line).  As
expected, the gap for protons, $\Delta_p$, and neutrons, $\Delta_n$,
solutions of the BCS system of two equations, in figure~\ref{bcs} are also
following similar variations, while the Fermi energies, $\lambda_p$ and
$\lambda_n$ have only a shallow minima in the surface region.  Deep minima
around $(R-R_i)/(R_t-R_i)=0.82$ are clearly seen in both figures.  Similar
results are also obtained for heavy nuclei like $^{240}$Pu,
figure~\ref{pu00lc} and $^{252}$Cf (see figure~\ref{cf00lc}).  See also
figures~1-3 of the e-print~\cite{p350ep16}.  At the touching point, $R=R_t$,
both kinds of variations of $R_2=R_2(R)$ are ariving at the same state,
hence the shell effects are identical there, as may be seen in
figures~\ref{shpe2}, \ref{pu00lc} and \ref{cf00lc}.

\begin{figure}
\centerline{\includegraphics[width=10cm]{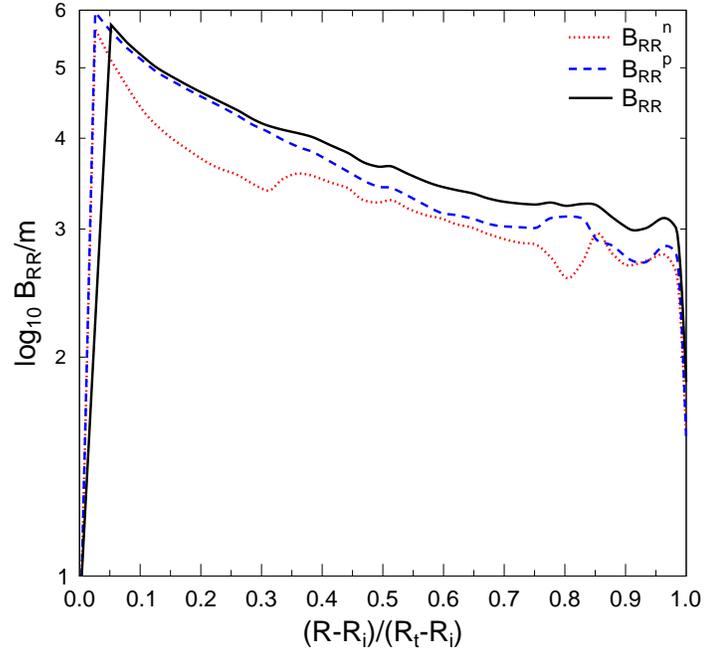}} 
\caption{(Color online) Decimal logarithm of the dimensionles RR component of
nuclear inertia tensor for symmetrical fission of $^{282}$Cn with linearly
increasing $R_2$.
\label{lgbr}} 
\end{figure}

The decimal logarithm of the dimensionles $B_{RR}/m$ component of nuclear
inertia tensor for symmetrical fission of $^{282}$Cn with linearly
increasing $R_2$ isplotted in figure~\ref{lgbr}.  At the touching point and
beyond, $R \geq R_t$, one should get the reduced mass: $B(R \geq R_t) =
mA_1A_2/A$.  The proton contribution are more important than the neutron
one.  This figure is completely different from the figure~5 of the
Ref.\cite{p333jpg14} where the components $B_{RR}/m$ for almost symmetrical
fission (with the light fragment $^{130}$Pd, $^{134}$Cd and $^{132}$Sn) of
$^{282}$Cn are shown for $R_2=$~constant.  Values larger than $10^5$ are
seen in figure~\ref{lgbr}, compared to smaller than $1.6 \times 10^3$ for
$R_2=$~constant.

If we use in graphics $x=(R-R_i)/(R_t-R_i)$ instead of $R$ 
then for $^{286}$Fl the interval of variation will be $x=(0, 1)$.
For the initial parent nucleus one may have either $x=0$ or/and $\eta =1$.
This is the reason why the dashed line ends up at the value of $\eta
=0.956$. In present calculations we have used 66 values of $x$ from $0$ to
$1.3$ and 24 values of $\eta$ from $0$ to $1$.

For minimization of action we need not only $B_{RR}$ but also the values of
$B_{R_2R_2}, B_{R_2R}$ 
in every point of a grid of 66$\times$24 for 66
values of $(R-R_i)/(R_t-R_i)$ and 24 values of $\eta = (A_1-A_2)/A$ or
$R_{2f}$.
We expect a dynamical path 
very different from the statical one shown in Fig.~\ref{cncl} with a white
dashed line.
The optimum value of the parameter zero-point vibration energy, $E_v$, used
to reproduce the experimental value of $^{282}$Cn spontaneous fission
half-life, $\log_{10} T_f^{exp} (s) =-3.086$.

In conclusion, with our method of calculating the spontaneous fission
half-life including macroscopic-microscopic method for deformation energy
based on asymmetric two-center shell model, and the cranking inertia for the
dynamical part, we may find a sequence of several trajectories one of which
gives the least action.  
Assuming spherical shapes, we found that the shape parametrization with
linearly increasing $R_2$ is more suitable to describe the fission process
of SHs in comparison with that of exponentially or linearly decreasing law. 
It is in agreement with the microscopic finding concerning the preformation
of a cluster at the surface, which then penetrates by quantum tunneling the
potential barrier.

\begin{acknowledgments} 

This work was supported within the IDEI Programme under Contracts No. 
43/05.10.2011 and 42/05.10.2011 with UEFISCDI, and NUCLEU Programme
PN16420101/2016 Bucharest.  

\end{acknowledgments}


\newpage

\newpage

\begin{table}[hbt] 
\caption{Statics.  Minima and maxima of deformation energy in MeV for
fission of $^{282}$Cn.  Linearly increasing $R_2$.  x$_{exit}$ corresponds
to $E_v=0.$ \label{tabl}}
\begin{center} \begin{ruledtabular} 
\begin{tabular}{c|cc|cc|cc|cc|c} $\eta$&x&1st min.&x&1st max.&x&2nd
min.  & x& 2nd max.  &x$_{exit}$\\ \hline 
0.000 & 0.000 & 0.000 &0.580 &29.114& 0.820 & 7.340 &0.900&14.720&1.15 \\ 
0.043 & 0.000 & 0.000 &0.580 &29.371& 0.840 & 7.211 &0.900&14.550&1.15 \\ 
0.087 &
0.000 & 0.000 &0.600 &29.515& 0.840 & 7.573 &0.920&15.045&1.25 \\ 0.130 &
0.000 & 0.000 &0.600 &29.817& 0.860 & 7.447 &0.940&14.884&1.27 \\ 0.174 &
0.000 & 0.000 &0.620 &29.911& 0.880 & 7.525 &0.960&15.033&1.34 \\ 0.217 &
0.000 & 0.000 &0.620 &30.172& 0.900 & 7.452 &0.980&15.853&1.40 \\ 0.261 &
0.000 & 0.000 &0.640 &30.267& 0.900 & 7.751 &1.000&15.488&1.44 \\ 0.304 &
0.000 & 0.000 &0.640 &30.090& 0.920 & 7.211 &0.980&17.500&1.45 \\ 0.348 &
0.000 & 0.000 &0.640 &30.327& 0.940 & 8.563 &1.040&15.755&1.46 \\ 0.391 &
0.000 & 0.000 &0.680 &30.585& 0.960 & 8.616 &0.980&10.345&1.34 \\ 0.435 &
0.000 & 0.000 &0.700 &30.647& 1.000 & 5.873 &1.000& 5.873&1.33 \\ 0.478 &
0.000 & 0.000 &0.720 &30.938& & & & & 1.46 \\ 0.522 & 0.000 & 0.000 &0.760
&29.417& & & & & 1.68 \\ 0.565 & 0.000 & 0.000 &0.860 &31.420& & & & & 1.73
\\ 0.609 & 0.000 & 0.000 &0.880 &31.611& & & & & 1.85 \\ 0.652 & 0.000 &
0.000 &1.180 &31.258& & & & & 2.06 \\ 0.696 & 0.000 & 0.000 &1.200 &34.233&
& & & & 2.39 \\ 0.739 & 0.000 & 0.000 &1.220 &33.409& & & & & 2.40 \\ 0.783
& 0.000 & 0.000 &1.240 &35.149& & & & & 2.80 \\ 0.826 & 0.000 & 0.000 &0.960
&40.492& & & & & 3.18 \\ 0.870 & 0.000 & 0.000 &0.960 &36.532& & & & & 4.23
\\ 0.913 & 0.000 & 0.000 &0.940 &35.736& & & & & 7.11 \\ 0.956 & 0.000 &
0.000 &0.940 &18.891& & & & &13.99 \\ \end{tabular} \end{ruledtabular}
\end{center} 
\end{table}
\begin{table}[hbt] 
\caption{Statics. Minima and maxima of deformation energy in MeV 
for fission of $^{282}$Cn. Constant $R_2$. 
x$_{exit}$ corresponds to $E_v=0.$
\label{tab}} 
\begin{center}
\begin{ruledtabular}
\begin{tabular}{c|cc|cc|cc|cc|c}
$\eta$ &x &1st min. &x &1st max. & x &2nd min. & x& 2nd max. &x$_{exit}$\\
\hline
0.000 & 0.100 & -1.458 &0.260 &0.737 & 0.420 &-1.480 &0.700 & 3.596 & 0.950 \\
0.043 & 0.100 & -1.323 &0.260 &1.390 & 0.440 &-1.910 &0.720 & 3.151 & 0.950 \\
0.087 & 0.100 & -1.114 &0.280 &2.102 & 0.460 &-2.061 &0.740 & 3.950 & 1.045 \\
0.130 & 0.100 & -0.920 &0.300 &2.845 & 0.500 &-1.779 &0.840 & 5.966 & 1.110 \\
0.174 & 0.100 & -0.715 &0.300 &3.639 & 0.520 &-1.061 &0.920 & 8.926 & 1.135 \\
0.217 & 0.100 & -0.529 &0.340 &4.467 & 0.540 & 0.029 &0.940 &11.298 & 1.165 \\
0.261 & 0.100 & -0.346 &0.360 &5.334 & 0.580 & 1.227 &0.960 &12.084 & 1.175 \\
0.304 & 0.100 & -0.226 &0.400 &5.796 & 0.600 & 2.502 &0.980 &13.256 & 1.190 \\
0.348 & 0.060 & -0.117 &0.420 &6.534 & 0.660 & 2.737 &0.980 & 9.722 & 1.170 \\
0.391 & 0.060 & -0.031 &0.440 &7.148 & 0.740 & 2.226 &1.000 & 4.888 & 1.150 \\
0.435 & 0.000 &  0.000 &0.480 &7.920 & 0.820 & 2.640 &1.040 & 4.173 & 1.150 \\
0.478 & 0.000 &  0.000 &0.540 &9.079 & 0.840 & 6.493 &1.060 & 8.094 & 1.190 \\
0.522 & 0.000 &  0.000 &1.060 &14.358& & & & & 1.670 \\
0.565 & 0.000 &  0.000 &1.080 &19.183& & & & & 1.696 \\
0.609 & 0.000 &  0.000 &1.080 &24.061& & & & & 1.644  \\
0.652 & 0.000 &  0.000 &1.100 &27.366& & & & & 2.107 \\
0.696 & 0.000 &  0.000 &1.100 &30.978& & & & & 2.051 \\
0.739 & 0.000 &  0.000 &1.120 &31.256& & & & & 2.447 \\
0.783 & 0.000 &  0.000 &1.140 &32.228& & & & & 2.461 \\
0.826 & 0.000 &  0.000 &1.180 &33.117& & & & & 3.454 \\
0.870 & 0.040 & -0.032 &1.220 &31.574& & & & & 3.952 \\
0.913 & 0.080 & -0.093 &1.280 &25.607& & & & & 6.933 \\
0.956 & 0.200 & -0.394 &1.808 &13.786& & & & & 8.266 \\
\end{tabular}
\end{ruledtabular}
\end{center}
\end{table}

\end{document}